\documentstyle[aps,prd,epsfig]{revtex}

\begin{document}
\preprint{SNUTP 04-005} \draft
\title{Stability of inflating branes in a texture}

\author{Inyong Cho\footnote{Electronic address:
iycho@phya.snu.ac.kr}}
\address{Center for Theoretical Physics, School of Physics,
Seoul National University, Seoul 151-747, Korea}

\date{\today}

\maketitle

\begin{abstract}
We investigate the stability of inflating branes embedded in an
O(2) texture formed in one extra dimension. The model contains two
3-branes of nonzero tension, and the extra dimension is compact.
When the gravitational perturbation is applied, the vacuum energy
which is responsible for inflation on the branes stabilizes the
branes if the symmetry-breaking scale of the texture is smaller
than some critical value. This critical value is determined by the
particle-hierarchy scale between the two branes, and is smaller
than the 5D Planck-mass scale. The scale of the vacuum energy can
be considerably low in providing the stability. This stability
story is very different from the flat-brane case which always
suffers from the instability due to the gravitational
perturbation.
\end{abstract}


\section{Introduction}
In ``braneworld'' models, our universe is represented by a
$(3+1)$-dimensional brane floating in a higher-dimensional bulk
spacetime~\cite{Rubakov,Arkani,RS}. The brane can be thought of as
fundamental $D$-brane, or it can arise as a topological defect in
a higher-dimensional field theory. In the simplest,
codimension-one models, the brane can be pictured as a domain wall
propagating in a 5D spacetime~\cite{Rubakov,Wall}. Higher
codimensions with both gauge and global defects have also been
considered.  In particular, models have been discussed where the
field configuration in the directions orthogonal to the brane is
that of a cosmic string~\cite{String} and a
monopole~\cite{Monopole}. In these models, the bulk curvature
produced by the defects plays a major role in localizing gravity
on the brane and in solving the mass-hierarchy problem.

Recently, the author investigated localization of gravity in a
different type of defect, the ``texture''~\cite{ChoTexture}.
Global textures are produced when a continuous global symmetry $G$
is broken to a group $H$. The resulting homotopy group
$\pi_D(G/H)$ is nontrivial, and the vacuum manifold is $S^D$; for
$G =\text{O}(N)$ models, $H=\text{O}(N-1)$ and $D=N-1$. The
mapping from the physical space to the vacuum manifold is $R^D \to
S^D$~\cite{Texture}.

For textures, the scalar field takes only the vacuum-expectation
value after the symmetry breaking, and thus nowhere in the
physical space remains the scalar field in the unbroken-symmetry
state. This is the key difference between textures and core
defects.

Let us consider a global O(2) texture formed in five dimensions.
In the mapping $R^1\to S^1$, we identify the physical space $R^1$
with the extra dimensional space. After the scalar field completes
its winding in the vacuum manifold, two spatial points on the
boundaries of the physical space are mapped into the same point in
the vacuum manifold. These two points are identified, and the
extra dimension becomes compact.

The extra dimension is curved by the texture. The strength of the
curvature depends on the symmetry-breaking scale of the texture.
This curved geometry of the extra dimension sets the particle
hierarchy as well as the Planck-mass hierarchy in the extra
dimension. Thus, the role of the cosmological constant in the
Randall-Sundrum scenario~\cite{RS} is replaced by the texture.

In Ref.~\cite{ChoTexture}, we investigated ``flat 3-branes''
embedded in a texture where the 4D worldsheet is represented by a
flat 4D metric. The model has two 3-branes of nonzero tension. The
one has a positive, and the other has a negative tension. The
particle-hierarchy scale between the two branes is determined
solely by the symmetry-breaking scale.
Gravity on the brane where TeV-scale particles are confined is
essentially four dimensional. The higher dimensional effect of
massive Kaluza-Klein gravitons is strongly suppressed on this
brane. This result is similar to that of Ref.~\cite{RSIII}.

One crucial problem of this model was that there exists a
tachyonic 4D graviton mode. Due to this mode, the direction
tangent to the brane is unstable although in the end it is
expected to relax to a final stable configuration after some
dynamical evolution combined with the texture field along the
orthogonal direction.

In this work, we are particularly interested in the mechanism to
stabilize the branes in gravitational perturbations. As a recipe,
we introduce ``inflation'' on 4D worldsheets.\footnote{The
inflating brane has been studied in different settings and
contexts, for example, in Refs.~\cite{BraneInf,Garriga}.} In doing
so, we effectively add vacuum energy induced by the bulk
curvature. When this energy compensates the energy deficit due to
the tachyon, we expect that the branes stabilize.

The scale of required vacuum energy in stabilizing the branes is
also of much concern. The expansion due to this energy should not
be very large to be consistent with the acceleration of the late
universe. We shall discuss this point also in this work.

The structure of this paper is as following. In
Sec.~\ref{sec=model}, we present the model, derive field
equations, and obtain the solutions. In Sec.~\ref{sec=graviton},
we introduce gravitational perturbations, and get the zero-mode
graviton. In Sec.~\ref{sec=check}, we discuss the stability
conditions in de Sitter space, and  check the stability of the
branes. In Sec.~\ref{sec=H}, we evaluate the expansion parameter
for stable branes, and we conclude in Sec.~\ref{sec=conclusion}.

\section{Field equations and solutions}\label{sec=model}
Let us consider an O(2) texture formed in one extra dimension.
The vacuum manifold is $S^1$, and it is mapped to $R^1$
which we set to be the extra dimensional space.
We shall consider a texture of one winding number.
We assume that the extra dimension is compact with the
$Z_2$ symmetry.
The two spatial points in the extra dimension, at which
the field takes the same value, are identified.
Then the orbifold is $S^1/Z_2$.

The action describing the model of an O(2) texture
and 3-branes in five dimensions is
\begin{equation}
{\cal S} = \int d^5x\sqrt{-g} \left[ {{\cal R}\over 16\pi G_N}
-{1\over 2}\partial_A\Phi^a \partial^A\Phi^a -{\lambda\over
4}(\Phi^a\Phi^a -\eta^2)^2 \right] -\int_i d^4x\sqrt{-h}
\sigma_i(\Phi^a) \,, \label{eq=action}
\end{equation}
where $\Phi^a$ is the scalar field with $a=1,2$,
$\eta$ is the symmetry-breaking scale,
$\sigma_i$ is the tension of the $i$-th brane, and
$g$ ($h$) is the 5D (4D) metric density.
For later convenience, we define
$\kappa^2 = 8\pi G_N = 1/M_*^3$, where
$M_*$ is the fundamental 5D Planck mass.

We adopt a metric ansatz in a conformal form,
\begin{equation}
ds^2 = g_{MN}dx^Mdx^N
= B(y)(d\bar{s}_4^2 +dy^2)\,,
\label{eq=metric}
\end{equation}
where $d\bar{s}_4^2$ is the 4D world-volume metric.

The scalar field resides in the vacuum manifold,
$\Phi^a\Phi^a=\eta^2$, and takes the form,
\begin{equation}
\Phi^a = \eta \left[ \cos\chi (y), \sin\chi (y) \right]\,,
\end{equation}
where $\chi (y)$ is the phase factor in the field space. In this
case, the nonlinear $\sigma$-model approximation is applied, and
the coupling $\lambda$ appearing in the action is treated as a
Lagrange multiplier. The scalar-field equation is then given by
\begin{equation}
\nabla^A\partial_A\Phi^a
=-{(\partial^A\Phi^b)(\partial_A\Phi^b) \over \eta^2}\Phi^a
+{\sqrt{-h} \over \sqrt{-g}}
{\partial\sigma_i(\Phi^a) \over \partial\Phi^a}\delta(y_i)
\,,
\end{equation}
which reduces to
\begin{equation}
\chi'' + {3\over 2}{B'\over B}\chi'
= {2\sqrt{B} \over \eta^2}\left[
{\partial\sigma_I \over \partial\chi}\delta(y)
+{\partial\sigma_{II} \over \partial\chi}\delta(y-y_*)
\right]\,.
\label{eq=sfd}
\end{equation}
Here, we assumed that there are two nonzero-tension branes located
at $y=0$ and $y=y_*$.

The energy-momentum tensor of the texture and the branes is given
by
\begin{equation}
T^M_N = \partial^M\Phi^a \partial_N\Phi^a -{1\over
2}\delta^M_N\partial_A\Phi^a\partial^A\Phi^a -{\sqrt{-h} \over
\sqrt{-g}} \sigma_i(\Phi^a)\delta (y_i) \text{diag} (1,1,1,1,0)
\,,
\end{equation}
then the Einstein's equation for the given metric and
energy-momentum tensor leads to
\begin{eqnarray}
-G^\mu_\mu &=& {1\over B} \left[ -{3\over 2}{B'' \over B} +{3\over
4}\left( {B'\over B} \right)^2 +{\bar{R}^{(4)} \over 4}\right] =
\kappa^2\eta^2 {\chi'^2 \over 2B} + {\kappa^2 \over \sqrt{B}}
\left[ \sigma_I(\Phi^a)\delta(y)
+ \sigma_{II}(\Phi^a)\delta(y-y_*) \right] \,,\\
-G^y_y &=& {1\over B} \left[ -{3\over 2}\left( {B'\over B}
\right)^2 + {\bar{R}^{(4)} \over 2}\right] =-\kappa^2\eta^2
{\chi'^2 \over 2B}\,.
\end{eqnarray}

Now, we consider inflation on 4D worldsheets, then the 4D
world-volume metric can be
\begin{equation}
d\bar{s}_4^2 = \bar{g}_{\mu\nu}dx^\mu dx^\nu = -dt^2 +
e^{2Ht}d\textbf{x}^2\,,
\end{equation}
and the 4D curvature scalar is $\bar{R}^{(4)} = 12H^2$. (Note what
is responsible for the expansion on the 4D worldsheet is not $H$,
but the ``effective''-expansion parameter $H_{eff}$ which will be
discussed in Sec.~\ref{sec=H}.)

The solutions to the Einstein and scalar-field equations are
\begin{eqnarray}
B(y) &=& \left\{ {B_0\over H}\text{sinh}[3H(|y|+y_0)]\right\}^{2/3}\,,
\label{eq=B}\\
\chi(y) &=& {\text{sign}(y)\over 3b_0}\ln \left[{\tanh {3H\over
2}(|y|+y_0) \over \tanh \left( {3Hy_0\over 2} \right)} \right]\,,
\label{eq=chi}
\end{eqnarray}
where $B_0=b_0\chi_0$, $b_0=\kappa\eta/2\sqrt{3}$, and $\chi_0$
and $y_0$ are integration constants. We take $B_0>0$ and $y_0>0$
without changing physics.

\begin{figure}
\begin{center}
\epsfig{file=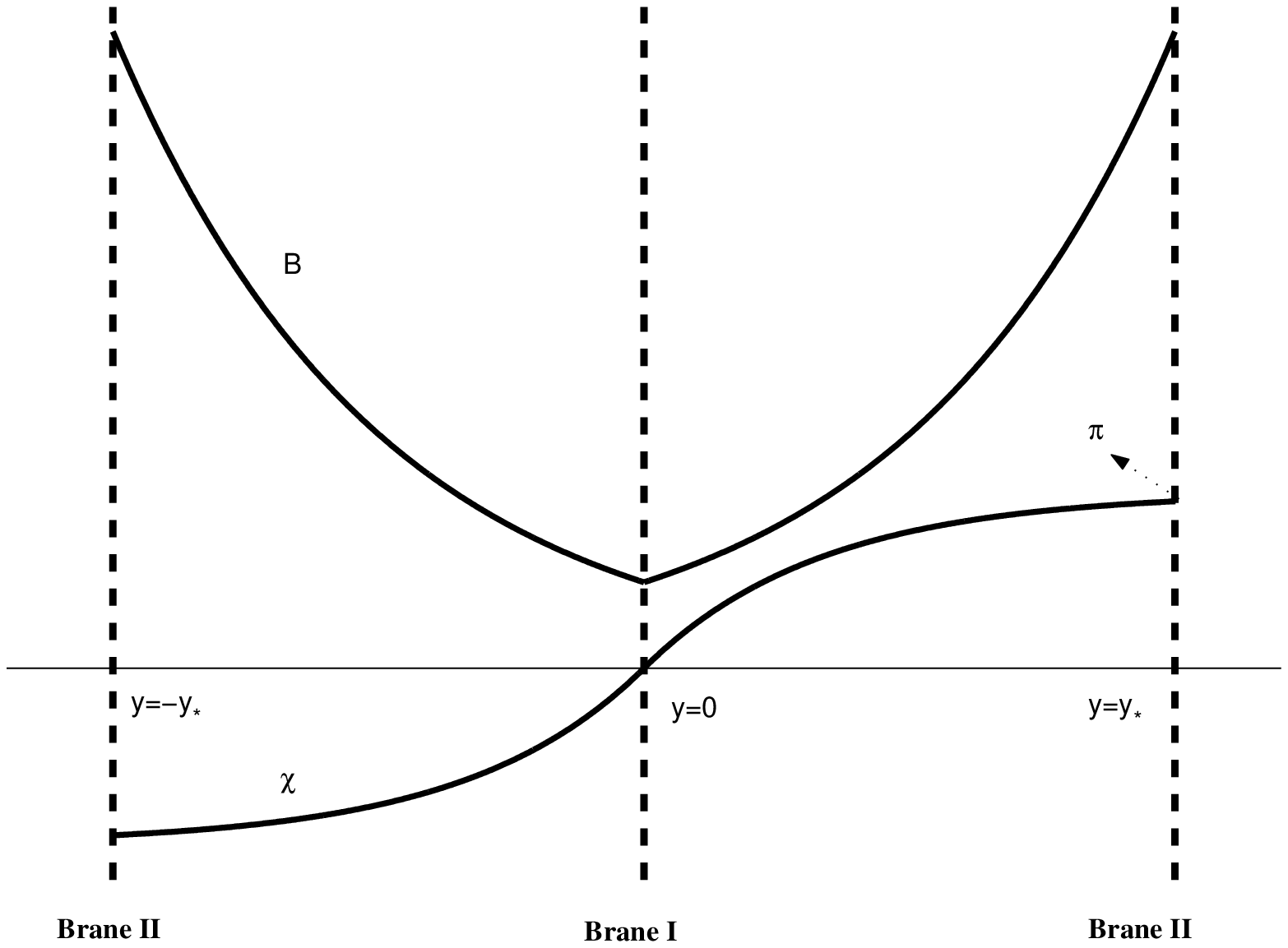,width=5in}
\end{center}
\vspace{0.5in} \caption{A schematic picture of the model and
solutions. Two branes are located at $y=0$ and $y=y_*$. Two
points, $y=y_*$ and $y=-y_*$ are identified, and the extra
dimension is compact. The gravitational field $B$ is lowest on the
first brane, and highest on the second brane. The scalar field
$\chi$ is continuous across the second brane because the $2\pi$
shift makes no change in $\chi$.} \label{fig=brane}
\end{figure}

The solutions are illustrated in Fig.~\ref{fig=brane}. The phase
factor $\chi$ varies from $-\pi$ to $\pi$ for the scalar field to
complete one winding in the vacuum manifold. We set $\chi(y=0)=0$
and $\chi (y=\pm y_*) =\pm\pi$. The boundaries at $y=\pm y_*$ are
identified since their phase factors correspond to the same point
in the vacuum manifold. Then this scalar-field configuration
preserves $Z_2$-symmetry about the branes.

From Eq.~(\ref{eq=chi}), setting boundary value $\chi (\pm y_*)=\pm\pi$
gives a condition,
\begin{equation}
{\tanh {3H\over 2}(y_*+y_0) \over \tanh\left({3Hy_0 \over 2}\right)}
=e^{3n}\,,
\label{eq=n}
\end{equation}
where we defined $n\equiv \pi b_0 = \pi\kappa\eta/2\sqrt{3}$.

If we plug the solutions (\ref{eq=B}) and (\ref{eq=chi}) into
the Einstein equations, we can evaluate the tension of the brane,
\begin{eqnarray}
\sigma_I &=& -{6H\over \kappa^2}
{1\over \tanh(3Hy_0)\left[{B_0\over H}\sinh(3Hy_0)\right]^{1/3}} <0\,,
\label{eq=sigmaI}\\
\sigma_{II} &=& {6H\over \kappa^2}
{1\over \tanh[3H(y_*+y_0)]\left[{B_0\over H}\sinh[3H(y_*+y_0)]\right]^{1/3}} >0\,.
\label{eq=sigmaII}
\end{eqnarray}
The first brane at $y=0$ has a negative tension, and the other has
a positive tension. This result is very similar to the flat-brane
case in Ref.~\cite{ChoTexture} as well as to the Randall-Sundrum
case~\cite{RS}. In addition, from the scalar-field equation
(\ref{eq=sfd}), we get
\begin{equation}
{d\sigma_I\over d\chi} ={d\sigma_{II}\over d\chi} =0
\end{equation}
at the location of the branes.

According to the warp factor $B(y)$, the particle-mass scale flows
along $y$. The mass scale is largest where the warp factor
acquires the largest value, i.e. on the second brane. We assume
that the Planck-scale particles are confined on the second brane
where the particle's effective mass is the same with the bare
mass.\footnote{Throughout this paper we assume that the particles
on the second brane are always of Planck scale while the particle
scale on the first brane varies depending on the situation we
consider.} Then, on the brane located at $y=y_i$, the
effective-particle mass measured relatively to the second-brane
particles is given by\footnote{For this derivation, please see
Refs.~\cite{RS,ChoTexture}.}
\begin{equation}
v_{eff} = \left\{ {\sinh[3H(y_i+y_0)] \over \sinh
[3H(y_*+y_0)]} \right\}^{1/3}v_i\,,
\label{eq=hier}
\end{equation}
where $v_i$ is the bare mass which we assumed Planck
scale.\footnote{Note that all the solutions and physical
properties obtained in this section and in the following sections
reduce to those of the flat-brane case in Ref.~\cite{ChoTexture}
in the limit of $H\to 0$.}

\section{Zero-mode graviton}\label{sec=graviton}
In this section, we investigate gravitational perturbations and
the zero-mode graviton. Let us consider a small perturbation
$h_{\mu\nu}$ to the background metric,
\begin{equation}
ds^2 = B(y)\left\{ \left[ \bar{g}_{\mu\nu} + {h_{\mu\nu}\over B(y)}
\right]dx^\mu dx^\nu +dy^2\right\}\,.
\label{eq=metrich}
\end{equation}
We assume that the only nonzero components are $h_{\mu\nu}$
($h_{\mu y} = h_{yy} =0$; the massive gravi-photon $h_{\mu y}$
decouples, and we assume no radion mode $h_{yy}$.), and apply a
transverse-traceless gauge on $h_{MN}$, $h_{MN|}{}^N=0$ and $h^M_M
=0$, where the vertical bar denotes the covariant derivative with
respect to the background metric. The Einstein's equation for
$h_{MN}$ reads
\begin{equation}
h_{MN|A}{}^A +2R^{(B)}_{MANB}h^{AB}
-2R^{(B)}_{A(M}h^A_{N)}
+{2(6-N)\over N-2}\kappa^2\sigma_i\delta (y_i)h_{MN}=0\,.
\label{eq=h}
\end{equation}
Here, the superscript $(B)$ represents that the quantity is
evaluated in the unperturbed background metric, and $N=5$ is the
number of the total dimensions (as distinguished from the index in
$h_{MN}$). We search the solution of a form, $h_{\mu\nu}
=h(y)\hat{e}_{\mu\nu}(x)$. In order to cast Eq.~(\ref{eq=h}) into
a nonrelativistic Schr\"odinger-type equation, we define a new
function by $\hat{h}(y) = B^{-1/4}h(y)$, then the equation leads
to
\begin{equation}
\left[ -{1\over 2}{d^2 \over dy^2} + V(y) \right]\hat{h}(y)
={m_g^2 \over 2}\hat{h}(y)\,, \label{eq=hhat}
\end{equation}
and the 4D part of the graviton satisfies the
equation~\cite{Garriga}
\begin{equation}
\Box^{(4)}\hat{e}_{\mu\nu} = (m_g^2+2H^2)\hat{e}_{\mu\nu}\,,
\end{equation}
where $m_g$ is the 4D graviton mass, and $\Box^{(4)}$ is the 4D
d'Alembertian evaluated in $\bar{g}_{\mu\nu}$. The potential is
given by
\begin{eqnarray}
V(y) &=&{9H^2\over 4}\left\{ 1-{1\over 2}\coth^2[3H(|y|+y_0)]\right\}
-\Sigma_I\delta(y)\,, \\
{}&=& {9H^2\over 4}\left\{ 1-{1\over
2}\coth^2[3H(-|y-y_*|+y_*+y_0)]\right\}
+\Sigma_{II}\delta(y-y_*)\,,
\end{eqnarray}
viewed from the brane I and II respectively, and
\begin{eqnarray}
\Sigma_I &=& {H\over 2}\coth(3Hy_0)\left\{1-4\left[{B_0\over H}\sinh (3Hy_0)
\right]^{1/3}\right\}\,,\\
\Sigma_{II} &=& {H\over 2}\coth[3H(y_*+y_0)]\left\{1-4\left[{B_0\over H}\sinh [3H(y_*+y_0)]
\right]^{1/3}\right\}\,.\\
\end{eqnarray}

For the zero mode, $m_g=0$, the solution to Eq.~(\ref{eq=hhat}) is
\begin{equation}
\hat{h}^I_0(y)=\sqrt{\sinh[3H(|y|+y_0)]}
\left\{a_1+a_2\ln\left[\tanh{3H\over 2}(|y|+y_0)\right]\right\}\,.
\label{eq=hI}
\end{equation}
Here, $a_1$ and $a_2$ are integration constants, and the
superscript $I$ means that the expression $|y|$ is valid on the
first brane. This zero mode represents a constant contribution
plus the contribution of the scalar field to the gravitational
perturbation in Eq.~(\ref{eq=metrich}). On the second brane, the
zero-mode wave function can be written as
\begin{equation}
\hat{h}^{II}_0(y)=\sqrt{\sinh[3H(-|y-y_*|+y_*+y_0)]}
\left\{a_1+a_2\ln\left[\tanh{3H\over 2}(-|y-y_*|+y_*+y_0)\right]\right\}\,.
\label{eq=hII}
\end{equation}
If we plug these solutions (\ref{eq=hI}) and (\ref{eq=hII}) in Eq.~(\ref{eq=hhat}),
we obtain two conditions at the boundaries, $y=0$ and $y=y_*$,
\begin{eqnarray}
-{a_1\over a_2} \equiv -a_{12} &=& {3\over
2\cosh(3Hy_0)\left\{1-\left[{B_0\over H}\sinh(3Hy_0)\right]^{1/3}\right\}}
+\ln\left[\tanh\left({3Hy_0\over 2}\right)\right]
\label{eq=a1a2I}\\
{}&=& {3\over
2\cosh[3H(y_*+y_0)]\left\{1-\left[{B_0\over H}\sinh[3H(y_*+y_0)]\right]^{1/3}\right\}}
+\ln\left[\tanh{3H\over 2}(y_*+y_0)\right]\,.
\label{eq=a1a2II}
\end{eqnarray}
These conditions are equivalent to imposing boundary conditions on
the branes, which are Sturm-Liouville type,
\begin{eqnarray}
\hat{h}_0(0) + \gamma_I \hat{h}_0'(0) &=& 0\,,\\
\hat{h}_0(y_*) + \gamma_{II} \hat{h}_0'(y*) &=& 0\,,
\end{eqnarray}
where
\begin{equation}
{1\over \gamma_i} = {H\over 2}\text{cotanh}[3H(y_i+y_0)]
\left\{1-4\left[{B_0\over
H}\sinh[3H(y_i+y_0)]\right]^{1/3}\right\}\,,
\end{equation}
and $y_i$ is the location of the brane, $y_I=0$ and $y_{II} = y_*$.

We can further determine $a_2$,
from the normalization condition of the zero-mode wave function,
$\int_0^{y_*} |\hat{h}_0(y)|^2dy= e_l/2$, where $e_l$ is the unit length.
To make the expression simpler, we introduce
$u_0\equiv 3Hy_0$ and $u_* \equiv 3H(y_*+y_0)$,
then from the normalization,
\begin{equation}
{e_l \over 2a_2^2}= {I_1 + I_2 +I_3 \over H}\,, \label{eq=a2}
\end{equation}
where
\begin{eqnarray}
{3\over a_{12}^2} I_1 &=& \cosh u_* -\cosh u_0\,,\\
{3\over 2a_{12}} I_2 &=& \cosh u_* \ln\left(\tanh {u_*\over
2}\right) -\cosh u_0 \ln\left(\tanh {u_0\over 2}\right) +u_* -u_0
-\ln\left( e^{2u_*}-1\right) +\ln\left( e^{2u_0}-1\right)\,,\\
{3\over 2} I_3 &=& \ln\left(\tanh {u_*\over 2}\right)\left[
{\tanh^2 {u_*\over 2} \over 1-\tanh^2{u_*\over
2}}\ln\left(\tanh{u_*\over 2}\right)
+\ln\left(1+\tanh {u_*\over 2}\right)\right]\nonumber\\
{}&-& \ln\left(\tanh {u_0\over 2}\right)\left[
{\tanh^2 {u_0\over 2} \over 1-\tanh^2{u_0\over 2}}\ln\left(\tanh{u_0\over 2}\right)
+\ln\left(1+\tanh {u_0\over 2}\right)\right] \\
{} &+& \int_{\tanh{u_*\over 2}}^{1+\tanh{u_*\over 2}} {\ln
(t)\over 1-t}dt -\int_{\tanh{u_0\over 2}}^{1+\tanh{u_0\over 2}}
{\ln (t)\over 1-t}dt\,.\nonumber
\end{eqnarray}

With the zero-mode wave function, the Newtonian gravitational
potential between the test masses $M_1$ and $M_2$ separated by the
distance $r$ on the brane at $y=y_i$, is given by
\begin{equation}
V_{Newt}(r) = G_N{|\hat{h}_0(y_i)|^2 \over \int |\hat{h}_0(y)|^2
dy} {M_1M_2 \over r}\,.
\end{equation}
The 4D gravitational constant on the first brane is then given by
\begin{eqnarray}
G_4(y=0) &=& G_N {|\hat{h}_0(0)|^2 \over e_l} \label{eq=G4a}\\
{} &=& G_N {a_2^2\over e_l}\sinh u_0 \left\{ {3\over 2\cosh u_0
\left[1-\left(
{B_0\over H}\sinh u_0\right)^{1/3}\right]}\right\}^2 \label{eq=G4b}\\
{} &=& G_N {H\sinh u_0\over 2(I_1+I_2+I_3)} \left\{ {3\over 2\cosh
u_0 \left[1-\left( {B_0\over H}\sinh
u_0\right)^{1/3}\right]}\right\}^2\,,\label{eq=G4}
\end{eqnarray}
where $G_N=1/8\pi M_*^3$, and if we assume that our universe is at
$y=0$ where we observe conventional 4D gravity, $G_4(0)=1/8\pi
M_{Pl}^2$.

\section{Stability check}
\label{sec=check} In the previous section, we explored the
gravitational perturbation and obtained the zero-mode graviton
wave function Remaining is the question whether the brane is
stable under this perturbation, or not. In the flat-brane
case~\cite{ChoTexture}, the model is unstable to the gravitational
perturbation; there exists one tachyonic mode in 4D gravitons,
which makes the direction tangent to the brane unstable. In the
inflating-brane model, we add energy on the brane in the form of
the cosmological constant. If this vacuum energy compensates the
energy deficit by the tachyonic graviton, the story of the
stability may be altered.

In flat space, the stability of the graviton is guaranteed if the
4D graviton mass squared is nonnegative, $m_g^2\geq 0$. However,
the stability condition in a de Sitter background is subtle and
needs some caution. The stability of massive high-spin fields has
been investigated very recently by Deser and Waldron~\cite{Deser}.
According to their work, in particular for the spin 2 fields
($s=2$) which represent gravitons, the stability story is as
following.

For the $m_g^2 =0$ mode, the graviton possesses two helicity
states, $\pm s = \pm 2$. This mode is always stable.

For $0<m_g^2 < 2H^2$, the graviton possesses full five helicity
states ($\pm 2$, $\pm 1$, 0), and the helicity-0 sector induces
instability. Therefore, the graviton in de Sitter space is
unstable in this regime, which is different from the flat-space
case.

For $m_g^2=2H^2$, so called the ``partially massless'' mode, the
dangerous helicity-0 sector disappears and the graviton is stable.

For $m_g^2>2H^2$, there exist full helicity states, but the
helicity-0 sector is not harmful, and the graviton is stable.

As a whole, in de Sitter space the graviton is stable if
$m_g^2=0$, or $m_g^2 \geq 2H^2$. Therefore, our policy in
searching stable solutions is twofold. As the graviton zero-mode
is always a solution satisfying the boundary conditions, first we
need to check if this zero mode is the lowest mode. This means
that there is no tachyonic mode, $m_g^2<0$, which is unstable.
Second, provided that the zero mode is the lowest mode, we check
if the first-excited massive graviton mode satisfies $m_g^2(1)
\geq 2H^2$ to avoid the unstable regime $0<m_g^2<2H^2$.

\subsection{Zero-mode check}
One simple way to examine if the zero mode is the lowest mode of
the wave equation, is to check the number of nodes of the
zero-mode wave function. For the Sturm-Liouville boundary value
problems, there exists a series of eigenvalues with a lower bound.
The eigenfunction corresponding to the lowest eigenvalue possesses
no node, and the number of nodes increases by one as the
eigenvalue increases to the next. If the zero-mode wave function
possesses no node between the boundaries, it will be the lowest
mode, and there will be no tachyonic mode.

If we look at the zero-mode wave function~(\ref{eq=hI}), it is a
monotonically increasing or decreasing function depending on the
constants. However, it is not clear yet if this function possesses
a node, or not. It depends on the value of constants in the wave
function. There are three constants to be determined, $y_0$,
$y_*$, and $B_0$ which we rescale as $u_0 \equiv 3Hy_0$, $u_*
\equiv 3Hy_*$ (as defined in the last section), and $\beta =
[(B_0/H)\sinh u_*]^{1/3}$.

In order to determine the above three constants, we use three
conditions introduced in the previous sections. First, we use the
boundary value of the scalar field, Eq.~(\ref{eq=n}),
\[
{\tanh (u_*/2) \over \tanh (u_0/2)} = e^{3n} \equiv \nu
\,,\qquad\text{(C.1)}
\]
where $n=\pi\kappa\eta/2\sqrt{3}$ is a free parameter that we can fix by
setting the symmetry-breaking scale.

The second condition comes from the particle hierarchy between the
two branes, Eq.~(\ref{eq=hier}),
\[
{v_{eff}(y=y_*) \over v_{eff}(y=0)} = \left({\sinh u_* \over \sinh
u_0} \right)^{1/3} \equiv 10^m \equiv \mu^{1/3}\,.
\qquad\text{(C.2)}
\]
Here, $m$ is a free parameter that we can fix by requiring the
particle hierarchy between the two branes. For example, if we
assume that the effective-particle mass on the first brane is of
TeV scale, we have $m=15$ and $\mu =10^{45}$. (As it was mentioned
earlier, the particle mass on the second brane is always set to
Planck scale.)

The third condition comes from the boundary conditions for the
zero-mode graviton, Eqs.~(\ref{eq=a1a2I}) and (\ref{eq=a1a2II}),
\begin{equation}
{1\over \cosh u_* (1-\beta)}-{1\over \cosh u_0
(1-\mu^{-1/3}\beta)} +2n=0 \,, \label{eq=third}
\end{equation}
which leads to a second-order polynomial equation for $\beta$,
\[
2n\mu^{-1/3}\beta^2 +\left[-2n(1+\mu^{-1/3})-{\mu^{-1/3}\over
\cosh u_*} +{1\over \cosh u_0}\right]\beta +2n+{1\over \cosh
u_*}-{1\over \cosh u_0} = 0\,. \qquad\text{(C.3)}
\]

To determine the constants, we solve (C.1) and (C.2), and get
\begin{equation}
\cosh u_0 = {1-2\nu/\mu +\nu^2 \over \nu^2 -1}\,,\quad
\cosh u_* = {1-2\mu\nu +\nu^2 \over 1- \nu^2}\,.
\end{equation}
As $\cosh u_i >1$, we get $\mu >\nu$. With these solutions for
$u_0$ and $u_*$, we can evaluate $\beta$ from (C.3). Then all the
constants are fully determined. In determining the constants, we
have two free parameters, $n$ and $m$. As was mentioned earlier,
$n$ is free related with the symmetry-breaking scale, and $m$ can
be fixed by the requirement of the particle hierarchy between the
two branes. We shall investigate the stability of the brane in a
various range of these parameters.

Now let us explore under what condition the zero-mode wave
function possesses no node in the region under consideration. Let
us denote the location of the node as $y=y_N$ if it existed. Then
the wave function satisfies $\hat{h}_0(y_N) = 0$. There will be no
node if $y_N>y_*$. We introduce $u_N \equiv 3H(y_N+y_0)$, then
from Eqs.~(\ref{eq=hI}) and (\ref{eq=a1a2I}) the condition
$\hat{h}_0(u_N)=0$ reduces to
\begin{eqnarray}
\tanh \left({u_N \over 2}\right) &=& \tanh \left( {u_0\over
2}\right) \exp\left[ {3 \over 2\cosh u_0 (1-\mu^{-1/3}\beta)}
\right]\label{eq=ineqL}\\
> \tanh \left({u_* \over 2}\right) &=& \tanh \left( {u_0\over
2}\right) \exp (3n)\,.\label{eq=ineqR}
\end{eqnarray}
Here, the inequality means that the node does not exist in the
range, $0\leq y \leq y_*$, i.e. $u_N>u_*$, and (C.1) was used for
Eq.~(\ref{eq=ineqR}). Keeping the no-node condition unchanged, the
inequality also holds in the same way for the exponents of
Eqs.~(\ref{eq=ineqL}) and (\ref{eq=ineqR}), and with the aid of
Eq.~(\ref{eq=third}), reduces to a simple form,
\begin{equation}
\cosh u_* (1-\beta) > 0\,.
\end{equation}
As a result, the condition for no-tachyon is simply
\begin{equation}
0<\beta <1\,. \label{eq=beta01}
\end{equation}

\begin{figure}
\begin{center}
\rotatebox{270}{\epsfig{file=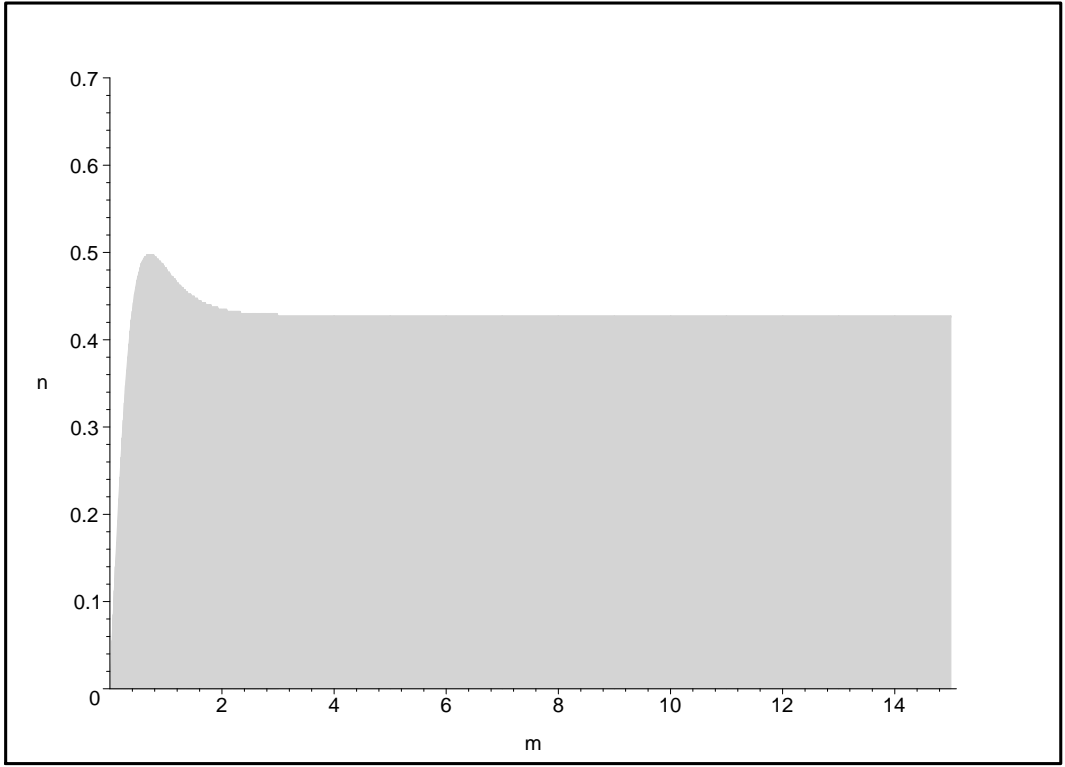,width=4in}}
\end{center}
\vspace{0.5in} \caption{The parameter space of $m$ vs. $n$. The
shaded region is where the graviton zero-mode is the lowest mode,
and the tachyonic-graviton mode is absent. $m=15$ corresponds to
the TeV hierarchy between the two branes, and $m=2$ corresponds to
the GUT hierarchy.} \label{fig=mvsn}
\end{figure}

We numerically search the values of the parameters $m$ and $n$ for
which the condition~(\ref{eq=beta01}) is satisfied. The resulting
domain in the parameter space $m$ vs. $n$ is plotted in
Fig.~\ref{fig=mvsn}. For a given $m$, there exists an upper bound
on $n$ under which the condition~(\ref{eq=beta01}) is satisfied.
For the most of the $m$ values, this condition is satisfied if
$n\lesssim 0.425$ ($\kappa\eta \lesssim 0.47$). The parameters
which give stable solutions will belong to a subset of this
domain.

\subsection{First-excited mode check}
Since the extra dimension of our model is compact, there exists a
discrete spectrum of massive Kaluza-Klein gravitons. In order to
satisfy the stability condition addressed earlier, the
first-excited massive mode should be $m_g^2(1) \geq 2H^2$. Let us
investigate the massive modes in this section.

The massive-mode solution to the wave equation (\ref{eq=hhat}) is
\begin{equation}
\hat{h}_{m_g}^I(y) = c_1 P_p^q[\coth [3H(|y|+y_0)]] + c_2
Q_p^q[\coth [3H(|y|+y_0)]]\,, \label{eq=hm}
\end{equation}
where $P$ ($Q$) is the associated Legendre function of the first
(second) kind, and
\begin{equation}
p=-{1\over 2}\,,\qquad q={1\over 2}\sqrt{1-{4m_g^2\over 9H^2}}\,.
\end{equation}
Similarly to the zero mode, $\hat{h}_{m_g}^{II}(y)$ is obtained by
making a change, $|y|+y_0 \to -|y-y_*|+y_*+y_0$, in
$\hat{h}_{m_g}^{I}(y)$, and from the boundary conditions at $y=0$
and $y=y_*$, we get
\begin{eqnarray}
-{c_1\over c_2} &=& {s_1Q_p^q(\coth u_0)
-(p+q)(p-q+1)Q_p^{q-1}(\coth u_0)
\over s_1P_p^q(\coth u_0) -(p+q)(p-q+1)P_p^{q-1}(\coth u_0)} \label{eq=c12a}\\
{}&=& {s_2Q_p^q(\coth u_*) -(p+q)(p-q+1)Q_p^{q-1}(\coth u_*) \over
s_2P_p^q(\coth u_*)-(p+q)(p-q+1)P_p^{q-1}(\coth u_*)}\,,
\label{eq=c12b}
\end{eqnarray}
where
\begin{eqnarray}
s_1 &=& \cosh u_0 \left[q + {1\over 6}(1-4\mu^{-1/3}\beta
)\right]\,,\\
s_2 &=& \cosh u_* \left[q + {1\over 6}(1-4\beta )\right]\,.
\end{eqnarray}
Equations~(\ref{eq=c12a}) and (\ref{eq=c12b}) give a mass
spectrum. It means that the massive mode exists only when the mass
satisfies this equality, so the spectrum is discrete, $m^2_g(i)$.

We numerically search the first-excited mode for given $m$ and $n$
which satisfies both the above mass-spectrum condition and
$m_g^2(1)\geq 2H^2$. First, in the domain of the $m$-$n$ plane
obtained in the previous section, let us fix $m$ and examine how
the value of $m_g^2(1)$ varies as $n$ varies. The numerical result
is given in Tab.~\ref{tab=m1} for $m=2$. If $n$ is large in the
domain, the value of $m_g^2(1)$ is smaller than $2H^2$, which
means unstable. As $n$ decreases, this value increases. Below some
critical value $n_c$, $m_g^2(1)>2H^2$ and the graviton becomes
stable. This behavior is observed also for larger $m$'s.

\begin{figure}
\begin{center}
\epsfig{file=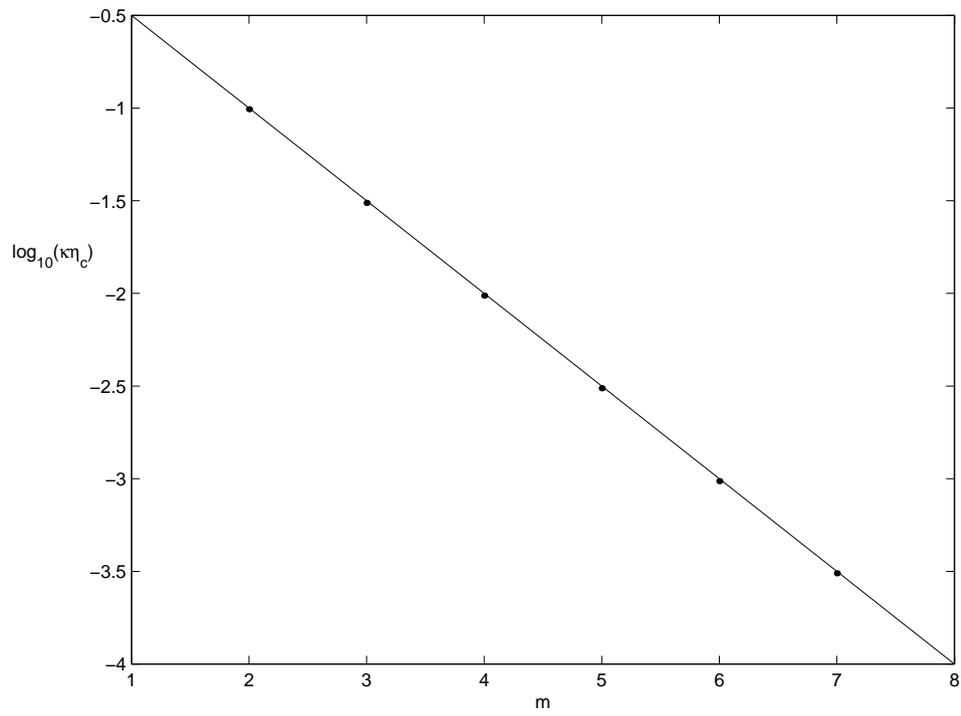,width=5in}
\end{center}
\vspace{0.5in} \caption{Plot of the critical value $\eta_c$ under
which there exist stable solutions. The numerical data (dots) fit
very accurately the reference function $-m/2$ (solid line), which
states $\kappa\eta_c \approx 10^{-m/2}$.} \label{fig=etac}
\end{figure}

Now we numerically search $n_c$ for several values of $m$. The
result is shown in Tab.~\ref{tab=nc} and plotted in
Fig.~\ref{fig=etac}. Amazingly, from Fig.~\ref{fig=etac} we
observe a very simple functional relation between $m$ and
$\kappa\eta_c \equiv 2\sqrt{3}n_c/\pi$, in a very high accuracy,
\begin{equation}
\kappa\eta_c = 10^{-m/2} = \sqrt{{v_{eff}(y=0) \over
v_{eff}(y=y_*)}}\,. \label{eq=etac}
\end{equation}
As $m$ increases, $\eta_c$ decreases exponentially, and the stable
domain in the $m$-$n$ plane gets smaller and smaller. For $m
\gtrsim 8$, the calculation goes beyond the present
numerical-resolution limit because the involved numbers are too
large, or small. However, we expect the above relation
(\ref{eq=etac}) continues for larger $m$'s.

We therefore conclude that there exists a domain in the $m$-$n$
plane in which the parameters give stable solutions. The
symmetry-breaking scale for stable solutions is very low, and the
upper bound is given by  Eq.~(\ref{eq=etac}). This result is very
different from the flat-brane case which was always
unstable~\cite{ChoTexture}.

\begin{table}
\begin{center}
\begin{tabular}{|c|c|c|c|c|c|c|}
  $n$ & 0.4 & 0.3 & 0.2 & 0.1 & 0.01 & 0.001 \\
  \hline
  $m_g^2(1)/H^2$ & 0.48991 & 1.35916 & 1.79189 & 1.98834 & 2.04406 & 2.04460 \\
\end{tabular}
\end{center}
\caption {The mass of the first-excited graviton for several
values of $n$, and for fixed $m=2$.} \label{tab=m1}
\end{table}

\begin{table}
\begin{center}
\begin{tabular}{|c|c|c|c|c|c|c|}
  $m$ & 2 & 3 & 4 & 5 & 6 & 7 \\
  \hline
  $n_c$ & 0.08939 & 0.02793 & 0.00882 & 0.00279 & 0.00088 & 0.00028 \\
\end{tabular}
\end{center}
\caption {The critical value $n_c$ for several values of $m$. }
\label{tab=nc}
\end{table}

\section{Effective-expansion parameter $H_{eff}$}\label{sec=H}

In this section, we evaluate the effective-expansion parameter
$H_{eff}$ for the stable branes. As we expected in the beginning,
adding vacuum energy to the brane can stabilize the texture-brane
system. The question is how much amount of the vacuum energy is
necessary, in other words, how large $H_{eff}$ should be to
stabilize the branes. Since $H_{eff}$ is responsible for the
expansion of the universe, it should be very small to be
consistent with the slow expansion of the late universe. The
current acceleration of the universe constrains the expansion
parameter as $H_{obs}\sim 10^{-42}\text{GeV}$.

From the 5D metric obtained in Sec.~\ref{sec=model}, the induced
4D metric on the 4D hypersurface at $y=y_i=\text{const}$ is given
by
\begin{eqnarray}
ds_4^2 &=& \left\{ {B_0\over H}\sinh[3H(|y_i|+y_0)]\right\}^{2/3}
(-dt^2+e^{2Ht}d\textbf{x}^2)\nonumber\\
{}&=& -d\tau^2 +e^{2H\left\{ {B_0\over
H}\sinh[3H(|y_i|+y_0)]\right\}^{-1/3}\tau}d\bar{\textbf{x}}^2\,,
\end{eqnarray}
where $\bar{\textbf{x}}$ is the rescaled 3D coordinates
$\textbf{x}$, and $\tau$ is the proper time measured on the
hypersurface at $y=y_i$,
\begin{equation}
\tau =\left\{{B_0\over H}\sinh[3H(|y_i|+y_0)]\right\}^{1/3}t\,.
\end{equation}
Then the 4D effective-expansion parameter is defined as
\begin{equation}
H_{eff} = H\left\{{B_0\over
H}\sinh[3H(|y_i|+y_0)]\right\}^{-1/3}\,.\label{eq=Heff}
\end{equation}
Here, for the given $m$ and $n$, $H$ is obtained from
Eq.~(\ref{eq=G4}),
\begin{equation}
{H\over M_*} = \left( {M_*\over M_{pl}}\right)^2
{2(I_1+I_2+I_3)\over \sinh u_0} \left[ {2\cosh u_0
(1-\mu^{-1/3}\beta) \over 3} \right]^2 \,. \label{eq=H}
\end{equation}

In determining $H_{eff}$ via $H$, we need to impose a constraint
on the fundamental 5D Planck mass, i.e. $M_{pl}/M_*$ in
Eq.~(\ref{eq=H}). First, we may assume that the 5D Planck-mass
scale is constrained in the same way as the particle scale on the
brane. That means for the given particle hierarchy between the two
branes, $v_{eff}(y_*)/v_{eff}(0) =10^m$, the same hierarchy
between the two Planck masses holds, $M_{pl}/M_* =10^m$. According
to this constraint, $M_* =\text{TeV}$ for $m=15$, and
$M_*=\text{GUT scale}$ ($=10^{16}\text{GeV}$) for $m=2$.

Second, regardless of the particle hierarchy, we fix the
fundamental Planck mass to TeV scale, $M_*=\text{TeV}$, so
$M_{pl}/M_* =10^{15}$.

Another constraint that we can consider in evaluating $H_{eff}$ is
the choice of the brane for residence. So far, we have assumed
that our universe is on the first brane. In this case, the
gravitational coupling is determined by, $G_4(y=0) = 1/8\pi
M_{pl}^2$, and from Eq.~(\ref{eq=Heff}) the 4D effective-expansion
parameter is given by
\begin{equation}
H_{eff} = {\mu^{1/3}\over \beta} H\,.
\end{equation}
However, if we assume that we are living on the second brane,
\begin{equation}
H_{eff} = {H\over \beta}\,.
\end{equation}
In this case, Eq.~(\ref{eq=G4a}) is rewritten as $G_4(y=y_*) =
1/8\pi M_{pl}^2 = G_N |\hat{h}(y_*)|^2/e_l$, the resulting $H$ is a
bit different from Eq.~(\ref{eq=H}), and is given by
\begin{equation}
{H\over M_*} = \left( {M_*\over M_{pl}}\right)^2
{2(I_1+I_2+I_3)\over \sinh u_*} \left[ {2\cosh u_* (1-\beta) \over
3} \right]^2 \,.
\end{equation}

\begin{figure}
\begin{center}
\rotatebox{270}{\epsfig{file=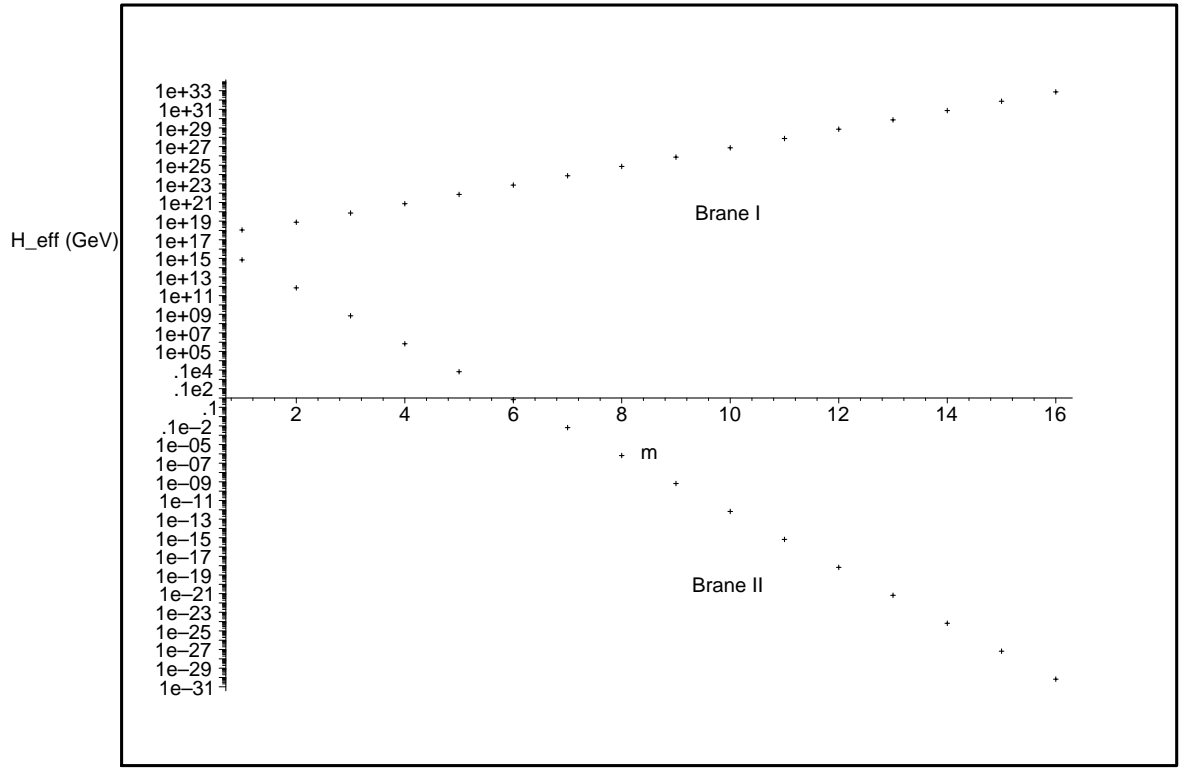,width=4in}}
\end{center}
\vspace{0.5in} \caption{ Plot of $H_{eff}$ vs. $m$ for $n=10^{-8}$
and $M_{pl}/M_*=10^m$. If we assume that our universe is on the
Brane I, $H_{eff}$ is very large $\protect\gtrsim
10^{19}\text{GeV}$. If we assume that our universe is on the Brane
II, $H_{eff}$ decreases as $m$ increases, and for the TeV
hierarchy ($m=15$), the expansion parameter can be lowered to
$H_{eff} \approx 10^{-27}\text{GeV}$. } \label{fig=Heff1}
\end{figure}

\begin{figure}
\begin{center}
\rotatebox{270}{\epsfig{file=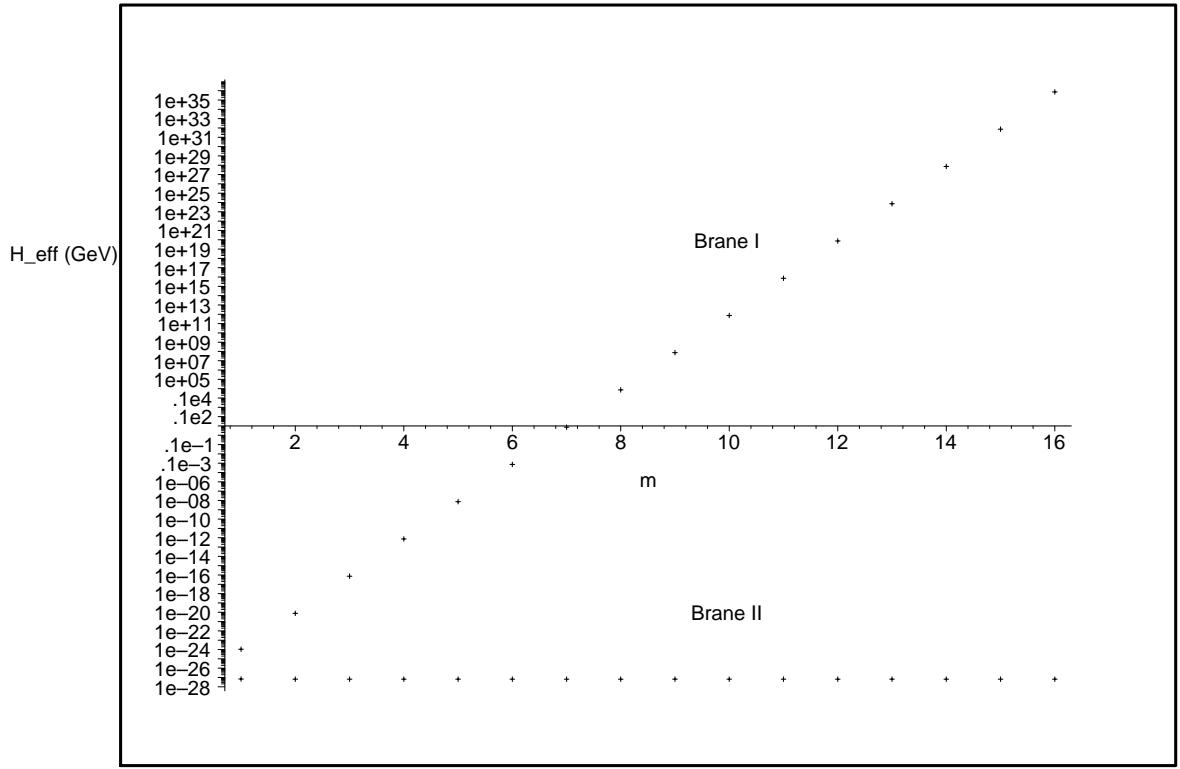,width=4in}}
\end{center}
\vspace{0.5in} \caption{A similar plot to Fig.~\ref{fig=Heff1},
but for assuming $M_*=\text{TeV}$. For the Brane I case, $H_{eff}$
decreases as $m$ decreases, and the expansion parameter can be
considerably lowered, $H_{eff}\approx 10^{-20}\text{GeV}$, for the
GUT hierarchy ($m=2$). For the Brane II case, $H_{eff}\approx
10^{-27}\text{GeV}$ regardless of $m$.} \label{fig=Heff2}
\end{figure}

Depending on the choices of the Planck-mass hierarchy $M_{pl}/M_*$
and the location of our universe in 5D, we have very different
$H_{eff}$ scales. The results are plotted in Fig.~\ref{fig=Heff1}
and Fig.~\ref{fig=Heff2}.

First, let us consider the case of $M_{pl}/M_*=10^m$.

(i) If we choose our universe is on the Brane I, i.e. $G_4(y=0) =
1/8\pi M_{pl}^2$, the 4D effective-expansion parameter is very
large $H_{eff}\gtrsim 10^{19}\text{GeV}$ regardless of the
particle-hierarchy scale $m$.

(ii) If we assume that our universe is on the Brane II (Planck
brane), i.e. $G_4(y=y_*) = 1/8\pi M_{pl}^2$, $H_{eff}$ decreases
as $m$ increases, and for the TeV hierarchy ($m=15$), the
expansion parameter can be lowered to $H_{eff}\approx
10^{-27}\text{GeV}$.

Second, let us consider that the fundamental 5D Planck mass is
fixed to TeV scale, $M_*=\text{TeV}$.

(i) For the Brane I case, $H_{eff}$ decreases as $m$ decreases,
and the expanding parameter can be considerably lowered,
$H_{eff}\approx 10^{-20}\text{GeV}$, for the GUT hierarchy, $m=2$.

(ii) For the Brane II case, $H_{eff}\approx 10^{-27}\text{GeV}$
regardless of $m$.

The numerical values of the results do not depend much on the
scale of $n$. In particular, for very small $n$'s, $H_{eff}$
remains almost unchanged as $n$ varies. The data in
Fig.~\ref{fig=Heff1} and Fig.~\ref{fig=Heff2} were obtained for
$n=10^{-8}$ which is supposed to give stable solutions for the
whole range of $m$.

\section{Conclusions}\label{sec=conclusion}
In this work, we investigated the gravitational perturbation on
the inflating 3-branes embedded in an O(2) texture formed in the
fifth dimension. In the previous work for the flat-brane
case~\cite{ChoTexture}, the gravitational perturbation made the
branes unstable. There existed one tachyonic mode in 4D gravitons.
The main purpose of this work was to investigate if the branes can
be stable to the perturbation when we allow the branes to inflate.

In a de Sitter background, the graviton is stable only if it is
massless, or if the mass ranges as $m_g^2 \geq 2H^2$. In the other
ranges, the graviton becomes unstable~\cite{Deser}. In this work,
we first checked if the zero-mode graviton of the texture-brane
solutions is the lowest mode to avoid $m_g^2<0$ modes. And then we
checked if the mass of first-excited mode is within the stable
range $m_g^2(1) \geq 2H^2$.

We found that there exist stable solutions if the
symmetry-breaking scale satisfies
\begin{equation}
\kappa\eta \lesssim \sqrt{{v_{eff}(y=0) \over
v_{eff}(y=y_*)}}\,,\label{eq=etac2}
\end{equation}
where $v_{eff}$ is the effective-particle mass on the brane. (We
assumed throughout this paper that the particles mass on the
second brane is of Planck scale, $v_{eff}(y=y_*) =
10^{18}\text{GeV}$.)

The scale of the effective-expansion parameter $H_{eff}$ for the
stable solutions varies depending on the constraints we impose.
Those constrains are the scale of the fundamental 5D Planck mass
$M_*$, the choice of the brane for our universe, and the
particle-hierarchy scale $m$ between the two branes.

In order to explain the very slow acceleration of the late
universe, we expect $H_{eff}$ to be very small. The possibly low
$H_{eff}$ scales are achieved mainly when we constrain the 5D
Planck mass to be $M_* =\text{TeV}$. The resulting $H_{eff}$
scales are, depending on the choices of the other constraints,

(i) $H_{eff}\approx 10^{-20}\text{GeV}$ : for assuming that our
universe is on the Brane I, and that the particle scale on the
Brane I is GUT scale.

(ii) $H_{eff}\approx 10^{-27}\text{GeV}$ : for assuming that our
universe is on the Brane II regardless of the particle scale on
the Brane I.

These $H_{eff}$ scales are still lager than the current expansion
parameter of the universe, $H_{obs}\sim 10^{-42}\text{GeV}$.
However, they are much smaller than the expansion rate during
inflation, $H_{inf}\sim 10^{10}\text{GeV}$. The stable
texture-brane system can be formed and survive at the late stages
in the evolution of the universe.

Once the symmetry-breaking scale $\eta$ is in the stable regime in
Eq.~(\ref{eq=etac2}), $H_{eff}$ is not very dependent on the
magnitude of $\eta$. Therefore, the values of $H_{eff}$ shown
above are globally valid for the whole stable texture-brane
systems, and we do not need to finely tune $\eta$ in order to have
those small values of $H_{eff}$.

\acknowledgements I am grateful to Jihad Mourad and Yoonbai Kim
for very useful comments and discussions. This work was supported
by Korea Science Foundation ABRL program (R14-2003-012-01002-0).

\clearpage

\clearpage

\clearpage

\clearpage

\clearpage

\end{document}